\documentclass[12pt]{article}

\usepackage{amsmath,stmaryrd}
\usepackage{amssymb,amsfonts,mathrsfs,bbm}

\usepackage{graphicx}
\usepackage{color}

\advance\voffset by -1.5cm
\advance\hoffset by -1.25cm
\newcommand{\tr}{{\rm tr}}
\newlength{\vs}\newlength{\vsh}
\numberwithin{equation}{section}

\begin{document}

\vspace*{-1.5cm}
\thispagestyle{empty}
\begin{flushright}
AEI-2009-075
\end{flushright}
\vspace*{2.5cm}

\begin{center}
{\Large
{\bf DBI analysis of generalised permutation branes}}
\vspace{2.5cm}

{\large Stefan Fredenhagen$^{1}$}
\footnotetext{$^{1}${\tt E-mail: stefan.fredenhagen@aei.mpg.de}}
{and \large Cosimo Restuccia$^{2}$}
\footnotetext{$^{2}${\tt E-mail: cosimo.restuccia@aei.mpg.de}}

\vspace*{0.5cm}

Max-Planck-Institut f{\"u}r Gravitationsphysik,
Albert-Einstein-Institut\\
D-14424 Golm, Germany\\
\vspace*{3cm}

{\bf Abstract}
\end{center}
We investigate D-branes on the product $G\times G$ of two group
manifolds described as Wess-Zumino-Novikov-Witten models. When the
levels of the two groups coincide, it is well known that there exist
permutation D-branes which are twisted by the automorphism exchanging
the two factors. When the levels are different, the D-brane charge
group demands that there should be generalisations of these
permutation D-branes, and a geometric construction for them was
proposed in hep-th/0509153. We give further evidence for this proposal
by showing that the generalised permutation D-branes satisfy the
Dirac-Born-Infeld equations of motion for arbitrary compact, simply
connected and simple Lie groups $G$.

\newpage

\section{Introduction}

In background geometries that contain the product of two identical
factors, $M\times M$, there exist so-called permutation branes which
are $\dim M$-dimensional submanifolds that lie diagonally in the
product $M\times M$. On the world-sheet of an open string, they are
described by permutation gluing conditions where the chiral
left-movers of one factor theory are glued to the corresponding
right-moving fields in the second factor and vice versa. When the
theory on $M$ is given by a rational conformal field theory (CFT), the
boundary states of the permutation branes can be explicitly
constructed~\cite{Recknagel:2002qq} (see
also~\cite{Figueroa-O'Farrill:2000ei,Gaberdiel:2002jr,Sarkissian:2003yw,Quella:2002ns}). 

It was observed in several instances that these permutation branes
play an important role for the D-brane charge group of the background
(see
e.g.\ \cite{Ashok:2004zb,Brunner:2005fv,Fredenhagen:2005an,Braun:2005eg}). In~\cite{Caviezel:2005th}
it was noted that in Gepner models a further construction is needed to
explain all the charges, some generalisation of the permutation branes
for products of $N=2$ minimal models with different central
charges. These have been constructed as matrix factorisations in a
Landau-Ginzburg description~\cite{Caviezel:2005th}, but no general
geometric or boundary conformal field theory construction of these
branes is known (see however~\cite{Fredenhagen:2006qw} for the CFT
construction in a particular product of minimal models). Earlier there
had been a proposal~\cite{Fredenhagen:2005an} for a generalisation of
permutation D-branes in the product of two Wess-Zumino-Novikov-Witten
(WZNW) models at different levels, $G_{k_{1}}\times G_{k_{2}}$. Also
there it was found that these generalisations are necessary to explain
the charge groups predicted by twisted topological K-theory.

In this note we want to further substantiate the proposal
of~\cite{Fredenhagen:2005an} by verifying that the generalised
permutation branes satisfy the Dirac-Born-Infeld (DBI) equations of
motion. Such a check has been performed in~\cite{Fredenhagen:2005an}
for the case of $G=SU (2)$, we shall extend it here to arbitrary
compact, simply-connected, simple Lie groups $G$.
\smallskip

Let us briefly review the geometry of generalised permutation branes
in $G_{k_{1}}\times G_{k_{2}}$ that was suggested
in~\cite{Fredenhagen:2005an}. Write $k_{i}=k_{i}'k$ where $k=\gcd
(k_{1},k_{2})$ is the greatest common divisor of the levels, so that
$k_{1}'$ and $k_{2}'$ are relatively prime. Then the simplest
generalised permutation brane\footnote{Notice that there are higher
dimensional cousins of~\eqref{genpermbrane}, which are not considered
in this note.} is described by the embedding
\begin{equation}\label{genpermbrane}
G \ni g \mapsto \big(g^{k_{2}'},g^{-k_{1}'} \big) \in G\times G \ .
\end{equation}
This embedding is motivated by the requirement that the H-field on the
brane has to be exact: as the H-field is proportional to the level,
the induced H-field has vanishing cohomology class on the brane.

We shall check in this paper that the D-brane described by the
embedding~\eqref{genpermbrane} together with the boundary two-form
given in~\eqref{boundarytwoform} (the gauge field on the brane)
satisfies the DBI equations of motion, and thus -- at least in the
geometric limit $k\to \infty$ -- defines a consistent D-brane.
\smallskip

The structure of the paper is as follows. In the remainder of the
introduction we shall set up the DBI equations of motion in the form
that is most useful for us. Section~2 introduces our choice of
coordinates on the groups. Finally, the actual computation will be
performed in section~3.

\subsection{DBI analysis}

The dynamics of geometric D-branes are described by the
Dirac-Born-Infeld (DBI)
theory~\cite{Fradkin:1985qd,Abouelsaood:1986gd,Callan:1986bc}. 
In the next paragraphs we shall review the DBI equations of motion,
following the formulation in~\cite{Ribault:2003sg}. Consider a brane
that is parameterised by coordinates $x^{a}$ with an embedding in the
target space (coordinates $X^{\mu}$) given by
\begin{equation}\label{embedding}
 x^{a} \mapsto X^{\mu}(x^a)\ .
\end{equation} 
In our case, the target space index runs over the two group factors
($\mu=1,\dots ,2d$, $d$ is the dimension of each group factor $G$ in the
target space), $x^a$ will run over the $d$ coordinates of the embedded
sub-manifold. We shall distinguish quantities on the brane from target
space objects by hats, so $\hat{g}$ and $\hat{B}$ are the induced
metric and B-field on the brane, respectively.

Our aim is to verify that the proposed embedding~(\ref{genpermbrane})
minimises the Dirac-Born-Infeld effective action,
\begin{equation}\label{DBI_action}
 S_{DBI}\propto \int d^{d}x\, \sqrt{\det (\hat{g}+\hat{B}+\hat{F})}\ ,
\end{equation} 
where $\hat{F}$ is the gauge field strength on the brane. The combined
object 
\begin{equation}
\hat{\omega} = \hat{B}+\hat{F}
\end{equation}
is the gauge-invariant boundary two-form. For the generalised
permutation brane~\eqref{genpermbrane} it has been proposed
in~\cite{Fredenhagen:2005an} to be (adapted to our normalisation)
\begin{align}
\hat{\omega} & = 
-\frac{k_{1}}{2}\sum_{j=1}^{k'_2-1}(k'_2-j)\,\tr\big(\text{Ad}_{g^j}(g^{-1}dg)\wedge
g^{-1}dg\big)\nonumber\\
&\quad - \frac{k_{2}}{2}\sum_{j=1}^{k'_1-1}(k'_1-j)\,
\tr\big(\text{Ad}_{g^{-j}}(g dg^{-1})\wedge g dg^{-1}\big)\ .
\label{boundarytwoform}
\end{align}
The DBI equations of motion are obtained from a variation of the DBI
action~\eqref{DBI_action}. In a gauge-invariant formulation they read
(see~\cite{Ribault:2003sg}),
\begin{align}\label{EOM} 
 \big[(\hat{g}+\hat{\omega})^{-1}\big]^{ba}\Omega^{\mu}_{ab} &=0\\
\label{F_field_EOM} 
 \partial_a \Big(
\sqrt{\det(\hat{g}+\hat{\omega})}\big[(\hat{g}+\hat{\omega})^{-1}\big]^{ab}_{\text{antisym}}\Big) &= 0 \ .
\end{align}
Here, $\Omega$ is defined to be a generalisation of the second
fundamental form, 
\begin{equation}\label{def_fF}
 \Omega^{\mu}_{ab}:=\partial_a\partial_b X^{\mu}+\Gamma'^{\mu}_{\nu\lambda}\partial_a X^{\nu}\partial_b X^{\lambda}-\hat{\Gamma}_{ab}'^c\partial_c X^{\mu},
\end{equation} 
with the connections entering this formula being defined in terms of
the H-field $H=dB$, 
\begin{equation}
 \Gamma':=\Gamma-\frac{1}{2} H\ .
\end{equation} 
The first equation~\eqref{EOM} corresponds to a variation of the
embedding, the second equation (\ref{F_field_EOM}) comes from the
variation of the F-field. For our problem at hand we shall compute the
generalised second fundamental form (\ref{def_fF}) explicitly in
section~3, with all the geometrical quantities involved. This will
then be used to check that the equations (\ref{EOM})
and (\ref{F_field_EOM}) are indeed satisfied for the generalised
permutation branes.\footnote{In~\cite{Fredenhagen:2005an} it was
already argued that~(\ref{F_field_EOM}) holds for arbitrary simple Lie
groups.  The argument given there, however, was not completely correct
and missed out a subtle point. See also footnote~\ref{fnote} on
page~\pageref{fnotepage}.} 
To turn the computation manageable we need to find good coordinates on
the brane and on the groups. This will be done in the next section.

\section{Coordinates on the group}

Our analysis starts by choosing a convenient parameterisation of the
single group manifold $G$. It was already suggested
in~\cite{Fredenhagen:2005an} that generalised permutation
branes are best described with a specific choice of coordinates. These
coordinates (described e.g.\ in~\cite{Maldacena:2001xj}) use the
triangular decomposition of the Lie algebra; every group
element is written as an element of a conjugacy class of an element of
the Cartan torus.

A convenient basis for a simple Lie algebra is given by the
Cartan-Weyl basis that consists of $r$ (the rank of the algebra)
commuting generators $H_{i}$ of the Cartan subalgebra, and the ladder
operators $E^{\alpha}$ associated to the roots $\alpha$. They satisfy
the commutation relations
\begin{align}
&  [H_i, E^{\alpha}]=\alpha_i
E^{\alpha} \ ,\qquad  [H_i,H_j]=0,\qquad i,j=1\dots r \\
\label{Cartan_Weyl_equations}
& [E^{\alpha},E^{\beta}] = \left\{ \begin{array}{ll}
\sum_{i}\alpha_{i} H_i & \textrm{if $\beta=-\alpha $}\\
 N_{\alpha\beta} E^{\alpha+\beta} & \textrm{if $\alpha+\beta\,\in \Delta$}\\
 0 & \textrm{otherwise}\\
  \end{array} \right.\ .
\end{align}
By $\Delta$ we denoted the set of all roots $\{\alpha\}$, $N_{\alpha
\beta}$ is some constant. We follow the usual
convention that the norm squared of a long root is $2$, and that
\begin{equation}\label{Killing_form}
\tr \big(H_i H_j\big)=\delta_{ij}\ ,\qquad \tr \big(E^{\alpha}
E^{\beta}\big)=\delta^{\alpha,-\beta}\ .
\end{equation} 
Now let us choose our
parameterisation. Following~\cite{Maldacena:2001xj} we write
\begin{equation}\label{parametrisation}
 g(\chi,\theta):=h^{-1}(\theta)t(\chi)h(\theta)\ ,
\end{equation} 
where $t$ is an element of the \textit{Cartan torus} $T$,
$t(\chi)=\exp(iH_j\chi^j)\,$, with $j$ running from $1$ to $r$, and
$h\in G$ is only defined up to left translations by $T$ (so it really
lives in the quotient space $G/T$).

One of the beautiful features of this parameterisation is that it
allows us to compute quite simply powers of a generic group element, 
\begin{equation}\label{g^n}
 g^n=h^{-1}t^nh\ .
\end{equation} 
This is particularly useful for the description of generalised
permutation branes, because the embedding~\eqref{genpermbrane}
involves powers of group elements. 

To complete the parameterisation we need in addition to the
coordinates $\chi$ on the Cartan torus to specify good coordinates on the
quotient space $G/T$. Following~\cite{Maldacena:2001xj} we first
introduce the one forms $\theta^{\alpha}$ by decomposing $dh\, h^{-1}$
in the Cartan-Weyl basis, 
\begin{equation}\label{theta_def}
 dh\, h^{-1}=\sum_{\alpha>0}i\big[\theta^{\alpha}E^\alpha+
\theta^{-\alpha}E^{-{\alpha}}\big]+i\zeta^i H_i \ .
\end{equation} 
Note that under a change of a representative $h$ of $G/T$, $h\to h'=f
h$ with $f =\exp (i \phi^{j}H_{j}) \in T$, the
decomposition~\eqref{theta_def} changes to
\begin{equation}
dh'\, h'^{-1}=\sum_{\alpha>0}i\big[e^{i\alpha \cdot \phi}\theta^{\alpha}E^\alpha+
e^{-i\alpha \cdot \phi} \theta^{-\alpha}E^{-{\alpha}}\big]+i
(\zeta^i+d\phi^{i}) H_i \ .
\end{equation}
So we see that the $\theta^{\alpha}$ only change by phases, whereas
the torus part $\zeta$ remains invariant only under constant rotations
$f$. Up to the problem with the phases, the one-forms $\theta^{\alpha}$
are well-defined objects on $G/T$, they will be used to locally
introduce coordinates on $G/T$.

We can express the metric and the H-field on the
group by the coordinates $\chi$ and the one-forms
$\theta^{\alpha}$. Following
again~\cite{Maldacena:2001xj,Fredenhagen:2005an}, on a
single group factor $G$ at level $k$ one finds
\begin{align}
  ds^2 & = -\frac{k}{2} \tr [g^{-1}dg\otimes_s g^{-1}dg] \\
\label{metric_generically}
  &=
\frac{k}{2}\Big[\sum_{\alpha>0}4 \sin^2{\frac{\chi^{i}\alpha_{i}}{2}}\big[\theta^{\alpha}\otimes
\theta^{-\alpha}+\theta^{-\alpha}\otimes\theta^{\alpha}\big]+\sum_{j=1}^{r}d\chi^j\otimes
d\chi^j \Big]\ .
\end{align}  
The phase ambiguity of the $\theta^{\alpha}$ obviously drops out,
because $\theta^{\alpha}$ and $\theta^{-\alpha}$ change by opposite
phases.
We can also express the H-field in this parameterisation,
\begin{equation}\label{B-field}
H = dB =
k d\Big[i\sum_{\alpha>0}\big(\chi^j\alpha_j-\sin{\chi^j\alpha_j}\big)\theta^{\alpha}\wedge
\theta^{-\alpha}\Big]\ .
\end{equation}  
Note that again the phase ambiguity in the $\theta^{\alpha}$ drops
out. As $H$ is not exact, the B-field defined by~\eqref{B-field} is
not globally well-defined, it is singular at those $t=\exp (iH\cdot
\chi)\in T$ different from the identity, whose conjugacy class
degenerates (and thus has dimension smaller than $G/T$).

It is instructive to see how the parameterisation in terms of $\chi$
and the $\theta$'s depend on the choice of a Cartan torus $T$. Suppose
we choose a different torus $T'$. By a well-known theorem, the two
tori are conjugate (see e.g.\ \cite[Chapter IV, Theorem
1.6]{Broecker_Dieck}), and there is a $g_{0}\in G$ with 
\begin{equation}
T' = g_{0}Tg_{0}^{-1}\ .
\end{equation}
The decomposition of a group element $g$ with respect to $T$ is then
related to the one with respect to $T'$ in a simple way,
\begin{equation}
g= h^{-1}t h = (g_{0}h)^{-1} g_{0}tg_{0}^{-1} (g_{0}h) = h'^{-1} t' h'\ ,
\end{equation}
where now $t'=g_{0}tg_{0}^{-1}$ and $h'=g_{0}h$. To find the
coordinates $\chi'$ and the one-forms $\theta'^{\alpha}$ we have to
use the changed basis of generators, $E'^{\alpha}=g_{0}E^{\alpha}g_{0}^{-1}$
and $H'_{i}=g_{0}H_{i}g_{0}^{-1}$. Then 
\begin{equation}
t' =g_{0}t g_{0}^{-1} = g_{0} e^{i\sum H_{j}\chi^{j}}g_{0}^{-1} =
e^{i\sum H'_{j}\chi^{j}} \ ,
\end{equation}
so that $\chi'^{j}= \chi^{j}$. On the other hand we have
\begin{equation}
dh' \, h'^{-1} = g_0(dh\, h^{-1})g_{0}^{-1} = \sum_{\alpha>0}
i\big[\theta^{\alpha}E'^\alpha+\theta^{-\alpha}E'^{-{\alpha}}\big]+i\zeta^i
H'_i \ ,
\end{equation}
so that $\theta'^{\alpha}=\theta^{\alpha}$. The parameterisation is
therefore independent of the choice of the Cartan torus.

We have seen that the parameterisation~\eqref{parametrisation} in
terms of $\chi$ and $\theta$ has very nice properties. On the other
hand we still face the problem that $\theta^{\alpha}$ are just
one-forms and not yet coordinates. If they were exact, we could
introduce coordinates by setting
$\theta^{\alpha}=dz^{\alpha}$. However, as we
shall see shortly, the one-forms $\theta^{\alpha}$ are not even
closed. Still it is possible to introduce coordinates $z$ locally
around $g=t (\chi)$ that capture all of the nice features and satisfy
\begin{equation}
\theta^{\alpha} =dz^{\alpha} + \mathcal{O} (z)\ .
\end{equation}
Namely we shall parameterise $h$ by
\begin{equation}\label{hparam}
 h = e^{i(\sum_{\alpha\in\Delta}z^{\alpha}E^{\alpha})}=1+i\sum_{\alpha\in\Delta}z_{\alpha}E^{\alpha}+\mathcal{O}(z^2)\;.
\end{equation} 
To get the one-forms $\theta^{\alpha}$ in terms of the variables $z$,
we expand $dh\, h^{-1}$ in $z$,
\begin{equation}
 dh\, h^{-1}=\sum_{\alpha,\beta\in\Delta^{+}}\Big[\Big(iE^{\alpha}dz^{\alpha}-\frac{1}{2}\big[E^{\alpha},E^{\beta}\big]dz^{\alpha}z^{\beta}\Big)+
\Big(iE^{-\alpha}d\bar{z}^{\alpha}-\frac{1}{2}\big[E^{-\alpha},E^{-\beta}\big]d\bar{z}^{\alpha}\bar{z}^{\beta}\Big)\Big]+\mathcal{O}(z^2)\ .
\end{equation}
By $\Delta^{+}\subset \Delta$ we denoted the set of all positive roots.
From this expression we can extract the one-forms $\theta^{\alpha}$,
$\theta^{-\alpha}$ by using~(\ref{theta_def})
and~(\ref{Killing_form}),
\begin{align}\label{theta_general}\begin{array}{l}
i\theta^{\alpha}=\tr \big(E^{-\alpha} dh\, h^{-1}\big)=i
dz^{\alpha}-\frac{1}{2}dz^{\beta}z^{\gamma}\,\tr \Big(E^{-\alpha}\big[E^{\beta},E^{\gamma}\big]\Big)+\mathcal{O}(z^2),\\ \\
i \theta^{-\alpha}=\tr \big(E^{\alpha}dh\, h^{-1}\big)=i
d\bar{z}^{\alpha}-\frac{1}{2}d\bar{z}^{\beta}\bar{z}^{\gamma}\,\tr \Big(E^{\alpha}\big[E^{-\beta},E^{-\gamma}\big]\Big)+\mathcal{O}(z^2),
\end{array}
\end{align}
We can further simplify the formulae (\ref{theta_general}) using the
Cartan-Weyl structure equations (\ref{Cartan_Weyl_equations}), and using the fact that $N_{-\alpha,-\beta}=-\overline{N}_{\alpha\beta}$,
\begin{align}\label{theta_general_2}
\theta^{\alpha}&= dz^{\alpha}+\frac{i}{2}\sum_{\substack{\beta,\gamma\in\Delta^{+}}}N_{\beta\gamma}dz^{\beta}z^{\gamma}\delta_{\beta+\gamma,\alpha}+\mathcal{O}(z^2)\\
 \theta^{-\alpha} & =
d\bar{z}^{\alpha}-\frac{i}{2}\sum_{\beta,\gamma\in\Delta^{+}}\overline{N}_{\beta\gamma}d\bar{z}^{\beta}\bar{z}^{\gamma}\delta_{\beta+\gamma,\alpha}+\mathcal{O}(z^2).
\end{align}
We see now explicitly that the $\theta^{\alpha}$ are in general not
closed, even if we sit on the Cartan torus ($z=0$),
\begin{align}
\label{dtheta1}
 d\theta^{\alpha}\rvert_{z,\bar{z}=0}
&=-\frac{i}{2}\sum_{\beta,\gamma\in\Delta^{+}}N_{\beta\gamma}\delta_{\beta+\gamma,\alpha}\,
dz^{\beta}\wedge dz^{\gamma}
\neq0 \quad  \text{in general}\\
 d\theta^{-\alpha}\rvert_{z,\bar{z}=0}
\label{dtheta2}
&=\frac{i}{2}\sum_{\beta,\gamma\in\Delta^{+}}\overline{N}_{\beta\gamma}\delta_{\beta+\gamma,\alpha}\,
d\bar{z}^{\beta}\wedge d\bar{z}^{\gamma}
\neq0 \quad  \text{in general.}
\end{align}
Expanding the metric (\ref{metric_generically}) in $z$ we find
\begin{align}\label{metric_single}
ds^2 &= \frac{k}{2} \Big[ \sum_{\alpha\in\Delta^+}
4\sin^2{\tfrac{\chi^{i}\alpha_{i}}{2}} \big[dz^{\alpha}\otimes
d\bar{z}^{\alpha} +d\bar{z}^{\alpha}\otimes dz^{\alpha}\big]
+\sum_{j=1}^{r}d\chi^j\otimes d\chi^j \Big] \nonumber\\ 
& \quad +\frac{k}{4} \Big[\sum_{\alpha,\beta ,\gamma >0}4
\sin^2{\tfrac{\chi^{i}\alpha_{i}}{2}} \delta_{\beta+\gamma,\alpha}
\big[iN_{\beta\gamma} z^{\gamma}\,(dz^{\beta}\otimes_s d\bar{z}^{\alpha})
-i\overline{N}_{\beta\gamma} \bar{z}^{\gamma}\,(dz^{\alpha}\otimes_s
d\bar{z}^{\beta})\big]\Big] \nonumber\\
&\quad +\mathcal{O}(z^2) \ .
\end{align} 
The first line of the previous expression is a real quadratic form; the second line involves a hermitian bilinear, as one can easily check.
The H-field on the other hand takes the following form:
\begin{align}
H &= k d\Big[i\sum_{\alpha>0}\big(\chi^j\alpha_j-\sin{\chi^j\alpha_j}\big)
dz^{\alpha}\wedge d\bar{z}^{\alpha}\Big] \nonumber\\
&\quad + \frac{ik}{2}d
\Big[\sum_{\alpha,\beta,\gamma >0}
\delta_{\beta+\gamma,\alpha} \big(\chi^j\alpha_j-
\sin{\chi^j\alpha_j}\big)
\,\big[iN_{\beta\gamma} z^{\gamma}\,(dz^{\beta} \wedge
d\bar{z}^{\alpha})-i\overline{N}_{\beta\gamma}
\bar{z}^{\gamma}\,(dz^{\alpha}\wedge d\bar{z}^{\beta})\big]\Big] \nonumber\\
&\quad + \mathcal{O}(z^2)\ .
\label{H_single}
\end{align}

\subsection{An example: coordinates on SU(2)}

We shall illustrate our choice of coordinates in the case of
$SU(2)$. Consider a standard parameterisation of $SU (2)$,
\begin{equation}
g(\psi ,\vartheta,\phi) =
\left( \begin{array}{cc}
\cos{\psi}+i\cos{\vartheta}\sin{\psi} & \sin{\psi}\sin{\vartheta}
e^{i\phi}  \\
-\sin{\psi}\sin{\vartheta} e^{-i\phi} &
\cos{\psi}-i\cos{\vartheta}\sin{\psi} \\
\end{array} \right) = h (\vartheta,\phi) t (\psi) h (\vartheta,\phi)^{-1},
\end{equation}
with
\begin{equation}
t(\psi) =
\left( \begin{array}{cc}
e^{i\psi} & 0 \\
0 & e^{-i\psi} \\
\end{array} \right)\ ,
\end{equation}
\begin{equation}
h(\vartheta,\phi) =
\left( \begin{array}{cc}
\cos{\frac{\vartheta}{2}} & -i\sin{\frac{\vartheta}{2}}e^{i\phi} \\[2mm]
-i\sin{\frac{\vartheta}{2}}e^{-i\phi} & \cos{\frac{\vartheta}{2}} 
\end{array} \right)\ .
\end{equation}
Using the standard form for the generators,
\begin{equation}
E^{+} =
\left( \begin{array}{cc}
0 & 1 \\
0 & 0 \\
\end{array} \right),\qquad
E^{-} =
\left( \begin{array}{cc}
0 & 0 \\
1 & 0 \\
\end{array} \right)\ ,
\end{equation}
we can extract the one-forms $\theta^{+}$ and $\theta^{-}$
from~\eqref{theta_def} and find
\begin{equation}\label{theta_su(2)}
 \theta^{+}=
-\frac{1}{2}e^{i\phi}d\vartheta-\frac{i}{2}\sin{\vartheta}e^{i\phi}d\phi,
\qquad\theta^{-}=\overline{\theta^{+}}\ .
\end{equation} 
Let us now identify the $z,\bar{z}$ variables in this simple
case. Following the definition~\eqref{hparam} we write
\begin{equation}
 h=e^{i(zE^++\bar{z}E^-)}= \left(\begin{array}{cc}
 \cos|z|&i\frac{z}{|z|} \sin|z|\\
i \frac{\bar{z}}{|z|}\sin|z|&\cos|z|
\end{array}
\right)\ .
\end{equation}
The connection of the coordinates $z,\bar{z}$ to the old
parameterisation is given by
\begin{equation}
 z=-\frac{\vartheta}{2}e^{i\phi} \ ,
\end{equation}
and the one-forms $\theta^{+},\theta^{-}$ (see~\eqref{theta_su(2)})
read now
\begin{align}
 \theta^{+} &=\Big(\frac{1}{2}+\frac{1}{4}\frac{\sin{2|z|}}{|z|}\Big)\,dz+\Big(\frac{1}{2}-\frac{1}{4}\frac{\sin{2|z|}}{|z|}\Big) \frac{z}{\bar{z}}\,d\bar{z} \nonumber\\
& = dz + \frac{1}{3}(z^2 d\bar{z}-|z|^2 dz)+\mathcal{O}(|z|^4)\ ,
\end{align}
and $\theta^{-}=\overline{\theta^{+}}$. These one-forms are not
closed, 
\begin{equation}
d\theta^{+} = \frac{\sin^{2}|z|}{\bar{z}} dz \wedge d\bar{z}\ ,
\end{equation}
but $d\theta^{+}$ vanishes at $z=0$. In contrast, for a generic Lie group,
$d\theta^{\alpha}$ does not have to vanish at $z=0$
(see~(\ref{dtheta1},\ref{dtheta2})). This is because in $SU (2)$ there are no
non-trivial relations among the positive roots, as there is only one.

\section{The DBI equations of motion}

We now want to check the DBI equations of motion. As the computations
are rather involved, we shall first outline the general strategy.

\subsection{The general strategy}

To check the equations of motion for the generalised permutation
branes at a given point of the brane parameterised by $g\in G$, we
choose a Cartan torus $T$ containing $g$. (This is always possible,
see e.g.\ \cite{Broecker_Dieck}.) Then we can introduce coordinates
$\chi^{j}$ and $z^{\alpha}$ locally around $g$ as in section~2. We
want to check the equations at $g$, that is at $z^{\alpha}=0$, but as
the equations of motion contain also first derivatives of the
geometric data (metric, B-field), we have to keep terms up to linear
order in $z$. The function describing the
embedding~\eqref{genpermbrane} enters even with second
derivatives, but it can be chosen such that it is exactly linear and
its second derivatives vanish,
\begin{equation}\label{embedding_coordinates}
(\chi,z) \mapsto \big((k_{2}'\chi,z),(-k_{1}'\chi
,z) \big)\ .
\end{equation}
This simple form of the embedding will in particular be very practical
when computing the connections.

The computations, although rather involved, simplify because of
two reasons: firstly, the target space is a direct product of two -- up to
the level -- identical factors, so that the target space data has a
natural block structure. Secondly, our choice of the coordinate system
allows us to factorise the geometry in a toroidal part (the directions
belonging to the Cartan torus), and a non-toroidal part. We can then
introduce ``matryoshka'' matrices, namely blocks of block matrices such
that many of the blocks are trivial and the computations can be
reduced considerably.

Before we start with the computation, let us introduce some notations
and conventions. For the brane we shall use the coordinates 
\begin{equation}\label{}
(x^{a})=(\chi^{i},z^{m},\bar{z}^{m})\ ,
\end{equation}
where the first $r=\text{rank}\, G$ coordinates $x^{a}$ denote the
toroidal parameters $\chi^{i}$ followed by the remaining $d-r$
coordinates corresponding to the $d-r$ roots. Here we introduced a
numbering of the positive roots $\alpha^{m}$ by a label $m$, and we
shall often write $z^{m}$ instead of $z^{\alpha^{m}}$.

Similarly, we parameterise the target space by coordinates $X^{\mu}$
where for $\mu =1,\dotsc ,d$ they denote the coordinates
$(\chi_{(1)}^{i},z_{(1)}^{m},\bar{z}_{(1)}^{m})$ of the first group
factor, and for $\mu =d+1,\dotsc ,2d$ they are given by the
coordinates of the second group factor. Hats $\hat{}$ denote
quantities on the brane, a tilde $\tilde{}$ denotes a geometric object
of a single factor of the product target space.

We divide the computations in several steps. First we determine the
expressions for the target space data in our coordinate system
(section~3.2). Then all quantities on the brane will be determined in
section~3.3, and in section~3.4 the second fundamental form will be
computed. In section~3.5 the equations~\eqref{F_field_EOM}
and~\eqref{EOM} will be checked showing that the generalised
permutation branes are extremal points of the DBI action.

\subsection{Target space Objects} 

The target space metric is just given by the
metrics~\eqref{metric_single} of the two group
factors,
\begin{align}\label{def_G}
 ds^2 & = \sum_{j=1,2} \Bigg\{ \frac{k_{j}}{2} \Big[\sum_{\alpha>0}
4\sin^2 \tfrac{\chi^{i}_{(j)}\alpha_{i}}{2} \big[dz_{(j)}^{\alpha}\otimes
d\bar{z}_{(j)}^{\alpha}+d\bar{z}_{(j)}^{\alpha}\otimes
dz^{\alpha}_{(j)}\big]
+\sum_{l=1}^r d\chi_{(j)}^{l}\otimes d\chi_{(j)}^{l} \Big]
\nonumber\\
& \quad + \frac{ik_{j}}{4} \Big[\sum_{\alpha,\beta ,\gamma >0}
4\sin^2{\tfrac{\chi^{i}_{(j)}\alpha_{i}}{2}}\, \delta_{\beta+\gamma,\alpha}
\big[N_{\beta\gamma}z_{(j)}^{\gamma}\, (dz_{(j)}^{\beta}\otimes_s
d\bar{z}_{(j)}^{\alpha})- \overline{N}_{\beta\gamma}\bar{z}_{(j)}^{\gamma}\,
(dz_{(j)}^{\alpha}\otimes_s d\bar{z}_{(j)}^{\beta})\big]\Big]\Bigg\}
\nonumber \\ 
& \quad +\mathcal{O} (z^{2})  \ .
\end{align}
Writing the metric $G_{\mu \nu}$ as a matrix it takes a block form,
\begin{align}\label{def_mat_G}
 \big( G(X^1,\dots, X^{2d})\big)_{\mu \nu}
=\frac{1}{2}\underbrace{\left(\begin{array}{c|}
k_1\tilde{G}(X^1,\dots, X^d)\vphantom{\widehat{G}} \\ \hline\hline
0\vphantom{\widehat{G^{\mu}}} 
\end{array}\right.\mspace{-1.5mu}}_d\underbrace{\mspace{-1.5mu}\left.\begin{array}{|c}
0\vphantom{\widehat{G^{\mu}}} \\ \hline\hline
k_2\tilde{G}(X^{d+1},\dots ,X^{2d})\vphantom{\widehat{G}} 
\end{array}\right)}_d\;.
\end{align}
The blocks of $d\times d$-matrices are separated by double lines. 
The constituent square matrix $\tilde{G}$ has again a
block structure with respect to toroidal and non-toroidal directions
(indicated by single lines), 
\begin{align}\label{def_Gtilde}
\tilde{G}(y,z) & =\left(\begin{array}{c|c}
\vphantom{\widehat{G}}\mathbbm{1}_{r} & 0  \\ \hline
0 &  \begin{array}{c|c}
\mathcal{O} (z^{2}) & \vphantom{\widehat{G}} \gamma(y,z)\\ \hline
\vphantom{\widehat{G}}\gamma(y,z)^{T} & \mathcal{O} (z^{2}) 
\end{array}\end{array}\right) 
\ .\\[-4mm]
& \qquad \underbrace{\hphantom{aa}}_{r}
\underbrace{\hphantom{aaaaaaaaaaaaa}}_{d-r} \mspace{40mu} \nonumber
\end{align}
The hermitian square matrix $\gamma$ has size $(d-r)/2$, and it is given by
\begin{equation}\label{def_gamma}
 \gamma_{mn}(y,z) = \left\{ \begin{array}{ll} 
4\sin^2\big( \tfrac{y^k\alpha^{n}_k}{2}\big)
+\mathcal{O}(z^{2})& \text{for}\ m=n\\[1mm]
2i\sin^{2}\frac{\chi^{j}\alpha_{j}^{n}}{2}
N_{\alpha^{m},\alpha^{n}-\alpha^{m}}z^{\alpha^{n}-\alpha^{m}} 
+\mathcal{O} (z^{2})
& \text{for}\ \alpha^{n}-\alpha^{m} \in \Delta^{+}\\[1mm]
-2i\sin^{2}\frac{\chi^{j}\alpha_{j}^{m}}{2}
\overline{N}_{\alpha^{n},\alpha^{m}-\alpha^{n}}\bar{z}^{\alpha^{m}-\alpha^{n}}
+\mathcal{O} (z^{2})
& \text{for}\ \alpha^{m}-\alpha^{n} \in \Delta^{+}\\[1mm]
\mathcal{O} (z^{2}) & \text{otherwise} \ .
\end{array}\right. 
\end{equation}
Notice that the diagonal does not have any linear
contribution in $z$. Furthermore, the off-diagonal terms are all at least
linear in $z$, but $\gamma_{mn}$ does neither depend on
$z^{\alpha^{m}}$ nor on $\bar{z}^{\alpha^{m}}$ linearly,
\begin{equation}\label{prop_gamma}
\partial_{a}\gamma_{mn}\Big|_{z=0} = 0 \quad \text{for}\ a\cong
z^{m},\bar{z}^{m}.
\end{equation}
The background H-field is also just the sum of the
H-fields~\eqref{H_single} of the two factors,
\begin{align}
 H &= \sum_{j=1,2} \frac{ik_j}{2}
d\Bigg[\sum_{\alpha>0}2\Big(\chi_{(j)}^{k}\alpha_k-\sin{\chi_{(j)}^{k}\alpha_k}\Big)dz_{(j)}^{\alpha}\wedge
d\bar{z}_{(j)}^{\alpha} \nonumber\\
&\quad + \sum_{\alpha,\beta,\gamma>0}\delta_{\beta+\gamma,\alpha}\Big(\chi_{(j)}^{k}\alpha_k-\sin{\chi_{(j)}^{k}\alpha_{k}}\Big)\,\big[iN_{\beta\gamma}z_{(j)}^{\gamma}\,(dz_{(j)}^{\beta}\wedge
d\bar{z}_{(j)}^{\alpha})-i\overline{N}_{\beta\gamma}\bar{z}_{(j)}^{\gamma}\,(dz_{(j)}^{\alpha}\wedge
d\bar{z}_{(j)}^{\beta})\big]\Bigg] \nonumber \\
&\quad + \mathcal{O} (z) .
\end{align}
Writing $H$ as $2d\times 2d$ matrices $H_{\mu}= (H_{\mu \nu
\lambda})_{\nu \lambda}$ we have again a block structure,
\begin{align}
H_{\mu} = \left\{ \begin{array}{ll}
 \dfrac{1}{2}\left(\begin{array}{c|}
\vphantom{\widehat{G}_{\mu}} k_1\tilde{H}_{\mu}(X^1\dots X^d)\\ \hline\hline
0
\end{array} \mspace{4mu}\begin{array}{|c}
\vphantom{\widehat{G}_{\mu}} 0\\ \hline\hline
0 
\end{array}\right) & \textrm{for $\mu=1,\dots ,d$}\\
\\
\dfrac{1}{2}\left(\begin{array}{c|}
0\\ \hline\hline
\vphantom{\widehat{G}_{\mu}}0
\end{array} \mspace{4mu}\begin{array}{|c}
0\\ \hline \hline 
\vphantom{\widehat{G}_{\mu}} k_2\tilde{H}_{\mu}(X^{d+1}\dots X^{2d})
\end{array}\right) & \textrm{for $\mu=d+1,\dots, 2d$.}
  \end{array} \right.
\end{align}
The form of the $d\times d$-matrices $\tilde{H}_{\mu}$ is different
for toroidal and non-toroidal directions $\mu$. For toroidal
directions we have 
\begin{equation}\label{def_Htilde_first_r}
 \big(\tilde{H}_{\chi^{i}}\big)_{ab}(y,z) =\left(\begin{array}{c|c}
0&0\\ \hline
0& \begin{array}{c|c}
\mathcal{O} (z)&\beta,_{i}(y,z)\\ \hline 
\vphantom{\widehat{G_{\mu}}}-\beta,_{i}^{T} (y,z)&\mathcal{O} (z)
\end{array}\end{array}\right) \ ,\quad \begin{array}{l}
 i=1\dots r\\
a,b=1\dots d
\end{array}
\end{equation}
Here we have introduced the anti-hermitian matrix $\beta$ of size $(d-r)/2$,
\begin{equation}\label{def_beta}
 \beta_{mn}(y,z)= \left\{ \begin{array}{ll}
2i\big( y^k\alpha^{n}_{k}-\sin{y^k\alpha^{n}_{k}}\big) +\mathcal{O}
(z^{2}) & \text{for}\ m=n \\[1mm]
-\big( y^k\alpha^{n}_{k}-\sin{y^k\alpha^{n}_{k}}\big)
N_{\alpha^{m},\alpha^{n}-\alpha^{m}}z^{\alpha^{n}-\alpha^{m}}
+\mathcal{O} (z^{2})
& \text{for}\ \alpha^{n}-\alpha^{m}\in \Delta^{+} \\[1mm]
\big( y^k\alpha^{m}_{k}-\sin{y^k\alpha^{m}_{k}}\big)
\overline{N}_{\alpha^{n},\alpha^{m}-\alpha^{n}}\bar{z}^{\alpha^{m}-\alpha^{n}}
+\mathcal{O} (z^{2})
& \text{for}\ \alpha^{m}-\alpha^{n}\in \Delta^{+}\\[1mm]
\mathcal{O} (z^{2}) & \text{otherwise} \ ,
\end{array}\right.
\end{equation} 
and its derivatives $\beta,_{i}(y,z):=\partial_i \beta(y,z)$. The matrix
$\beta$, similarly to $\gamma$, has linear terms in $z$ only in
off-diagonal terms, and it enjoys a analogous property
to~\eqref{prop_gamma},
\begin{equation}\label{prop_beta}
\partial_{a}\beta_{mn}\big|_{z=0} = 0 \quad \text{for}\ a\cong
z^{m},\bar{z}^{m}.
\end{equation}
For the non-toroidal directions it will be
sufficient for the computation to observe that $\tilde{H}$ has the form
\begin{equation}\label{def_Htilde_non_toroidal}
 \big(\tilde{H}_{z^{m}}\big)_{ab}(y,z) = 
\left(\begin{array}{c|c}
0& *\\ \hline 
* & \begin{array}{c|c}
* & \boxbslash \\ \hline
\boxbslash & *
\end{array}
\end{array}\right) 
+ \mathcal{O} (z) \qquad (\text{similarly for}\ \tilde{H}_{\bar{z}^{m}})\ ,
\end{equation}
where $*$ denotes an arbitrary contribution, whereas $\boxbslash$
denotes a matrix that has vanishing diagonal, $\boxbslash_{ij}=0$ for
$i=j$.

The only target space data that is still missing for the computation
is the connection,
\begin{equation}
 \Gamma_{\mu\lambda\nu}:=\frac{1}{2}\Big(\partial_{\lambda}G_{\mu\nu}+\partial_{\nu}G_{\mu\lambda}-\partial_{\mu}G_{\nu\lambda}\Big)\ .
\end{equation}  
Again we write it in matrix form where it assumes a
block structure,
\begin{equation}
(\Gamma_{\mu})_{\nu \lambda} = \left\{ \begin{array}{ll}
 \dfrac{1}{2}\left(\begin{array}{c|}
\vphantom{\widehat{G}_{\mu}} k_1\tilde{\Gamma}_{\mu}(X^1\dots X^d)\\
\hline\hline 0
\end{array} \mspace{4mu}\begin{array}{|c} \vphantom{\widehat{G}_{\mu}}
0\\
\hline \hline 0 \end{array}\right) & \textrm{for $\mu=1,\dots ,d$}\\[6mm]
  \dfrac{1}{2}\left(\begin{array}{c|} 0 \\
\hline\hline \vphantom{\widehat{G}_{\mu}} 0 
\end{array} \mspace{4mu}\begin{array}{|c}
0\\ \hline\hline
\vphantom{\widehat{G}_{\mu}} k_2\tilde{\Gamma}_{\mu}(X^{d+1}\dots X^{2d}) 
\end{array}\right) & \textrm{for $\mu=d+1,\dots 2d$}
  \end{array} \right.
\end{equation}
with
\begin{align}
 \big(\tilde{\Gamma}_{\chi^{i}}\big)_{ab}(y,z) &=-\frac{1}{2}\left(\begin{array}{c|c}
0&0\\ \hline
0&\begin{array}{c|c}
0&\gamma,_{i}(y,z)\\ \hline 
\gamma,_{i}(y,z)&0
\end{array}\end{array}\right) +\mathcal{O} (z) ,\quad \begin{array}{l}
 i=1\dots r\\
a,b=1\dots d
\end{array}\\
\label{def_christ_non_toroidal}
\big(\tilde{\Gamma}_{z^{m}}\big)_{ab} (y,z) &=
\left(\begin{array}{c|c}
0& *\\ \hline 
* & \begin{array}{c|c}
0 & \boxbslash \\ \hline
\boxbslash & 0
\end{array}
\end{array}\right) 
+\mathcal{O} (z)\qquad (\text{similarly for}\ \tilde{\Gamma}_{\bar{z}^{m}})\ , 
\end{align} 
where as before $*$ is not further specified, and $\boxbslash$ is a matrix
with vanishing diagonal terms. 

\subsection{Brane quantities}

The quantities on the world-volume of the brane that are important in
our analysis are first of all the induced metric $\hat{g}$ and the
boundary two-form $\hat{\omega}$, which we have to determine up to
linear order in $z$, and then their derivatives, namely the connection
$\hat{\Gamma}$ and the induced H-field $\hat{H}$, for which we only
need the expressions at $z=0$.

We start with the induced metric, determined by
\begin{equation}
 \hat{g}_{ab}:=G_{\mu\nu}\partial_a X^{\mu}\partial_b X^{\nu}\ .
\end{equation} 
The contributions from the two factors add up, and we find
\begin{equation}\label{expl_induced_metric}
\big( \hat{g}\big)_{ab} (\chi,z)  = \left(\begin{array}{c|c}
\frac{k_{1}'k_{2}' (k_{1}+k_{2})}{2} \mathbbm{1}_r&0\\[1mm] \hline
0&\frac{1}{2}\left(\begin{array}{c|c}
\mathcal{O} (z^{2})&\hat{\gamma} (\chi,z) \\ \hline 
\hat{\gamma}^{T} (\chi,z)&\mathcal{O} (z^{2})
\end{array}\right)
\end{array}\right) \ ,
\end{equation}
where the matrix $\hat{\gamma}$ of size $(d-r)/2$ is given
in terms of $\gamma$ (defined in~\eqref{def_gamma}),
\begin{equation}
\hat{\gamma} (\chi,z)  = k_{1}\gamma(k_2'\chi,z) + k_{2}\gamma(-k_1'\chi,z)\ .
\end{equation}
The connection $\hat{\Gamma}$ corresponding to the induced metric
$\hat{g}$, 
\begin{equation}
 \hat{\Gamma}_{cab}:=\frac{1}{2}\Big(\partial_{b}\hat{g}_{ca}+\partial_{a}\hat{g}_{cb}-\partial_{c}\hat{g}_{ab}\Big)\;,
\end{equation}  
is then obtained as
\begin{align}
\Big(\hat{\Gamma}_{\chi^{i}}\Big)_{ab} &=-\frac{kk_{1}'k_{2}'}{4}
\left(\begin{array}{c|c}
0&0\\ \hline
0&\begin{array}{c|c}
0&\gamma,_{i}(k_2'\chi)-\gamma,_{i}(-k_1'\chi)\\ \hline 
\gamma,_{i}(k_2'\chi)-\gamma,_{i}(-k_1'\chi)&0\\
\end{array}
\end{array}\right) + \mathcal{O} (z)\\
\Big(\hat{\Gamma}_{z^{m}}\Big)_{ab} &= 
\left(\begin{array}{c|c}
0& *\\ \hline 
* & \begin{array}{c|c}
0 & \boxbslash \\ \hline
\boxbslash & 0
\end{array}
\end{array}\right)  + \mathcal{O} (z) \qquad (\text{similarly for}\
\hat{\Gamma}_{\bar{z}^{m}})\ .
\end{align}
In addition to the metric data, we also have to specify the gauge
field living on the brane, which is specified by the boundary two-form
$\hat{\omega}$ given in~\eqref{boundarytwoform}. It is not
difficult to show that in terms of $\chi$ and $\theta$ it can be expressed as 
\begin{equation}
\hat{\omega}= i\sum_{\alpha>0}\big(k_{2}\sin{k_{1}'\chi^j\alpha_j}-k_{1}\sin{k_{2}'\chi^j\alpha_j}\big)\theta^{\alpha}\wedge
\theta^{-\alpha} \ .
\end{equation}
Expressed in the local coordinates $(\chi,z)$ it has the form
\begin{equation}\label{expl_induced_B}
\big(\hat{\omega}\big)_{ab} (\chi,z) =
\left(\begin{array}{c|c}
0&0\\ \hline
0&\frac{1}{2}\left(\begin{array}{c|c}
\mathcal{O} (z^{2})&\hat{\beta} (\chi,z)\\ \hline 
-\hat{\beta}^{T} (\chi,z)&\mathcal{O} (z^{2})
\end{array}\right)
\end{array}\right) \ ,
\end{equation}
where $\hat{\beta}$ is expressed by $\beta$ (defined
in~\eqref{def_beta}),
\begin{equation}
\hat{\beta} (\chi,z)  =  k_{1} \beta (k_{2}'\chi,z) 
+ k_{2}\beta(-k_{1}'\chi,z)\ .
\end{equation}
The exterior derivative of the boundary two-form gives the induced
H-field, $\hat{H}=d\hat{\omega}$,
\begin{align}
\Big(\hat{H}_{\chi^{i}}\Big)_{ab} &= \frac{kk_{1}'k_{2}'}{2}
\left(\begin{array}{c|c}
0&0\\ \hline
0&\begin{array}{c|c}
0&\beta,_{i}(k_2'\chi)-\beta,_{i}(-k_1'\chi)\\ \hline 
-\beta,_{i}(k_2'\chi)+\beta,_{i}(-k_1'\chi)&0\\
\end{array}
\end{array}\right) + \mathcal{O} (z) \\
\Big(\hat{H}_{z^{m}}\Big)_{ab} &= 
\left(\begin{array}{c|c}
0& * \\ \hline 
* & \begin{array}{c|c}
* & \boxbslash\\ \hline
\boxbslash & *
\end{array}
\end{array}\right) + \mathcal{O} (z) 
\qquad (\text{similarly for}\ \hat{H}_{\bar{z}^{m}})\ .
\end{align}
Now we have all geometric data at our disposal to compute the
(generalised) second fundamental form that appears in the DBI
equations of motion.

\subsection{Generalised second fundamental form}

The generalisation of the second fundamental form $\Omega_{ab}^{\mu}$
in the presence of a background H-field is given in~(\ref{def_fF}). As
our embedding~\eqref{embedding_coordinates} is linear, the second
derivatives $\partial_{a}\partial_{b}X^{\mu}$ vanish, and
$\Omega_{ab}^{\mu}$ is just
\begin{equation}
 \Omega^{\mu}_{ab} = \Gamma'^{\mu}_{\nu\lambda}\partial_a
X^{\nu}\partial_b X^{\lambda}-\hat{\Gamma}_{ab}'^c\partial_c X^{\mu} 
=:\Omega^{(TS)\mu}_{ab}-\Omega^{(WS)\mu}_{ab}\ .
\end{equation} 
In the following, we shall need the combinations
\begin{equation}\label{def_gg}
\mathscr{G}^{\pm}_{i}(y) = \big( \gamma,_{y^{i}}(y,z)\pm
\beta,_{y^{i}}(y,z)\big)_{z=0} \ .
\end{equation}
If $\mu$ corresponds to a torus direction, the quantities
$\Omega^{(TS)\mu}$ and $\Omega^{(WS)\mu}$ have relatively simple expressions,
\begin{align}
\Big(\Omega^{(TS)\mu}\Big)_{ab}&=\Big(\Gamma^{\mu}_{\lambda\nu}-\frac{1}{2} H^{\mu}_{\lambda\nu}\Big)\partial_a X^{\lambda}\partial_b X^{\nu}=
G^{\rho\mu}\Big(\Gamma_{\rho\lambda\nu}-\frac{1}{2}
H_{\rho\lambda\nu}\Big)\partial_a X^{\lambda}\partial_b X^{\nu} \nonumber\\
&=\left\{ \begin{array}{ll}
-\dfrac{1}{2} \left(\begin{array}{c|c}
0&0\\ \hline
0&\begin{array}{c|c}
0& \mathscr{G}^{+}_{i} (k_{2}'\chi) \\ \hline 
\vphantom{\widehat{A}_{\mu}} (\mathscr{G}^{-}_{i})^{T} (k_{2}'\chi) &0
\end{array}
\end{array}\right) + \mathcal{O}(z) & \text{for}\ \mu
\cong \chi_{(1)}^{i} \\[8mm] 
-\dfrac{1}{2} \left(\begin{array}{c|c}
0&0\\ \hline 
0&\begin{array}{c|c} 0&  \mathscr{G}^{+}_{i} (-k_{1}'\chi) \\
\hline \vphantom{\widehat{A}_{\mu}}  (\mathscr{G}^{-}_{i})^{T} (-k_{1}'\chi) &0
\end{array}
\end{array}\right) + \mathcal{O}(z) & \text{for}\ \mu \cong
\chi_{(2)}^{i}\end{array}
\right.
\end{align}
and
\begin{align}\label{1_tor_WS}
 \Big(\Omega^{(WS)\mu}\Big)_{ab}&=\Big(\hat{\Gamma}^{c}_{ab}-\frac{1}{2}\hat{H}^{c}_{ab}\Big)\partial_c X^{\mu}=\Big(\hat{\Gamma}_{nab}-\frac{1}{2}\hat{H}_{nab}\Big)\hat{g}^{nc} \partial_c X^{\mu}\nonumber\\
&\mspace{-36mu} = \left\{ \begin{array}{r}
-\dfrac{1}{2}\dfrac{k_2}{k_1+k_2} 
\left(\begin{array}{c|c}
0&0\\ \hline
0&\begin{array}{c|c}\vphantom{\widehat{A}_{\mu}}
0&\mathscr{G}^{+}_{i}(k_2'\chi)-\mathscr{G}^{+}_{i}(-k_1'\chi) \\ \hline \vphantom{\widehat{A}_{\mu}}
(\mathscr{G}^{-}_{i})^{T}(k_2'\chi)-(\mathscr{G}^{-}_{i})^{T}(-k_1'\chi)&0\\
\end{array}
\end{array}\right)  \\[7mm]
+\mathcal{O}(z) \  \text{for} \ \mu \cong
\chi_{(1)}^{i}\\[8mm]
\dfrac{1}{2}\dfrac{k_1}{k_1+k_2} 
\left(\begin{array}{c|c}
0&0\\ \hline
0&\begin{array}{c|c}\vphantom{\widehat{A}_{\mu}}
0&\mathscr{G}^{+}_{i}(k_2'\chi)-\mathscr{G}^{+}_{i}(-k_1'\chi) \\ \hline\vphantom{\widehat{A}_{\mu}} 
(\mathscr{G}^{-}_{i})^{T}(k_2'\chi)-(\mathscr{G}^{-}_{i})^{T}(-k_1'\chi)&0\\
\end{array}
\end{array}\right) \\[7mm]
+\mathcal{O}(z)\  \text{for}\ \mu \cong
\chi_{(2)}^{i} \end{array}\right.
\end{align}
The expressions for $\Omega^{\mu}$ for non-toroidal directions $\mu$
are more involved, we however only need the block structure,
\begin{equation}
\Omega^{\mu} = \left( \begin{array}{c|c}
0 & * \\ \hline
* & \begin{array}{c|c}
* & \boxbslash \\ \hline
\boxbslash & *
\end{array}
\end{array}\right) + \mathcal{O}(z) \quad \text{for}\ \mu \cong
z^{m},\bar{z}^{m} \ .
\end{equation}

\subsection{Checking the equations of motion}

Now we have all the elements to check that the proposed geometry
(\ref{genpermbrane}) satisfies the DBI equations of motion (\ref{EOM})
and (\ref{F_field_EOM}). An important quantity in the DBI equations is
the sum $\hat{g}+\hat{\omega}$.  From the equations
(\ref{expl_induced_metric}) and (\ref{expl_induced_B}) we find
\begin{equation}\label{expl_gplusomega}
\big(\hat{g}+\hat{\omega}\big)_{ab}=
\left(\begin{array}{c|c}
\big[\frac{k_{1}'k_{2}'}{2}(k_1+k_2)\big]\mathbbm{1}_r&0\\[1mm] \hline
0&\frac{1}{2}\left(\begin{array}{c|c}
\mathcal{O} (z^{2})& \hat{\gamma}+\hat{\beta}\\ \hline
\vphantom{\overline{\widehat{G^{\beta}}}}\overline{\hat{\gamma} + \hat{\beta}} & \mathcal{O} (z^{2}) \\
\end{array}\right)
\end{array}\right)\ .
\end{equation}
The explicit expression for the constituent matrices is
\begin{equation}\label{expl_gammaplusbeta}
\frac{1}{2}\big(\hat{\gamma}+ \hat{\beta}\big)_{mn} =  
\left\{\begin{array}{ll}
k_{1}+k_{2}-\big(k_1 e^{ ik_2'(\chi^k\alpha^{n}_k)}+k_2
e^{- ik_1'(\chi^k\alpha^{n}_k)}\big)+ \mathcal{O}
(z^{2})& \text{for}\ m=n\\[2mm]
\mathcal{O} (z) & \text{for} \ m\not= n \ .
\end{array} \right.
\end{equation}
Notice again that there are no linear terms in $z$ on the
diagonal, and that the off-diagonal terms are all at least of linear
order with the property
\begin{equation}\label{prop_gammaplusbeta}
\partial_{a} \big(\hat{\gamma}+\hat{\beta} \big)_{mn}\Big|_{z=0} =0
\quad \text{for}\ a\cong z^{m},\bar{z}^{m}\ ,
\end{equation}
which follows from~\eqref{prop_gamma} and~\eqref{prop_beta}.
The inverse of $(\hat{g}+\hat{\omega})$ that enters the DBI equations
of motion is given by
\begin{equation}\label{gplusomega_inverse}
\Big( \big(\hat{g}+\hat{\omega}\big)^{-1}\Big)^{ab} = 
\left(\begin{array}{c|c}
\Big[\frac{2}{k_{1}'k_{2}'(k_1+k_2)}\Big]\mathbbm{1}_r&0\\[2mm] \hline
0& 2\left(\begin{array}{c|c}
0 & \Big(
\overline{\hat{\gamma}+\hat{\beta}}\Big)^{-1\vphantom{G^{G}}}\\[2mm] 
\hline 
\Big( \hat{\gamma} + \hat{\beta} \Big)^{-1\vphantom{G^{G}}} & 0 
\end{array}\right) + \mathcal{O} (z^{2})
\end{array}\right)\ .
\end{equation}
\smallskip

We are now prepared to check the equations of motion. Let us start
with the equation coming from the variation of the gauge field.

\subsubsection{Gauge field equation of motion}

Let us recall the gauge field equation of motion
(see~\eqref{F_field_EOM}),
\begin{equation}\label{F_field_EOMr}
 \partial_a \Big(
\sqrt{\det(\hat{g}+\hat{\omega})}\big[(\hat{g}+\hat{\omega})^{-1}\big]^{ab}_{\text{antisym}}
\Big)=0\ .
\end{equation}
From the explicit form~\eqref{gplusomega_inverse} of
$\big( \hat{g}+\hat{\omega}\big)^{-1}$, we see that for toroidal 
directions, $b \cong \chi^{i}$, we have
\begin{equation}
\big[(\hat{g}+\hat{\omega})^{-1}\big]^{a \chi^{i}}_{\text{antisym}} =
0 \ ,
\end{equation}
so that the equation of motion~\eqref{F_field_EOMr} is satisfied for
toroidal directions $b$.

For non-toroidal directions $b$, only derivatives in non-toroidal
directions $a$ occur in the equation of
motion~\eqref{F_field_EOMr}. As it suffices to check the equation at
$z=0$, we have to investigate the linear terms in $z$ in the
expression in parentheses in~\eqref{F_field_EOMr}. For the determinant
of $\hat{g}+\hat{\omega}$ we have
\begin{equation}
\det (\hat{g}+\hat{\omega}) \propto |\det (\hat{\gamma
}+\hat{\beta})|^{2} + \mathcal{O} (z^{2})\ . 
\end{equation}
Now the matrix structure of $\hat{\gamma}+\hat{\beta}$ comes into
play, which is of the form (see~\eqref{expl_gammaplusbeta}),
\begin{equation}\label{formofgammaplusbeta}
\hat{\gamma}+\hat{\beta} = D + N (z) + \mathcal{O} (z^{2})\ ,
\end{equation}
where $D$ is a diagonal matrix independent of $z$ and $N (z)$ is
off-diagonal and linear in $z$. This means that the determinant
of $\hat{\gamma}+\hat{\beta}$ has no linear term -- in the Leibniz sum
formula for the determinant there is no summand that contains
precisely one off-diagonal matrix element -- and hence its
derivative with respect to any $z^{m}$ vanishes at $z=0$. 

It remains to discuss the linear terms
in~$\big[(\hat{g}+\hat{\omega})^{-1}\big]$. Its building blocks
(see~\eqref{gplusomega_inverse}) are
given by the inverse of~\eqref{formofgammaplusbeta},
\begin{equation}
\big( \hat{\gamma}+\hat{\beta}\big)^{-1} = D^{-1} - D^{-1}N (z)D^{-1} 
+ \mathcal{O} (z^{2})\ .
\end{equation}
The linear terms in $( \hat{\gamma}+\hat{\beta})^{-1}$ 
are all off-diagonal, and to leading order the off-diagonal elements
are proportional to the off-diagonal elements of
$\hat{\gamma}+\hat{\beta}$, so that the
property~\eqref{prop_gammaplusbeta} carries over 
to~$(\hat{\gamma}+\hat{\beta})^{-1}$, 
\begin{equation}
\partial_{a} \Big(\big(\hat{\gamma}+\hat{\beta} 
\big)^{-1}\Big)^{mn}\bigg|_{z=0} =0
\quad \text{for}\ a\cong z^{m},\bar{z}^{m}.
\end{equation}
From this we can finally 
conclude\label{fnotepage}\footnote{\label{fnote}The gauge
field equations of motion were also investigated
in~\cite{Fredenhagen:2005an}, and for the toroidal directions the
correct argument was already given there. For the non-toroidal
directions, however, it was implicitly assumed that already the matrix
$\hat{\gamma}+\hat{\omega}$ has no linear terms in $z$ which is in
general not correct.} that the equation of motion~\eqref{F_field_EOMr}
is also satisfied for non-toroidal directions $b$.

\subsubsection{Embedding equations of motion}

It remains to check the equation of motion~\eqref{EOM} that comes from
the variation of the embedding of the brane,
\begin{equation}\label{EOMr} 
 \big[(\hat{g}+\hat{\omega})^{-1}\big]^{ba}\Omega^{\mu}_{ab}=0\ .
\end{equation}
For non-toroidal directions $\mu$, the block structure of the matrices
that are involved is enough to verify the equation of motion,
\begin{equation}
\tr \big( (\hat{g}+\hat{\omega})^{-1} \Omega^{\mu} \big)\Big|_{z=0} = 
\tr \left(\left(\begin{array}{c|c}
* & 0 \\ \hline
0 & \begin{array}{c|c}
0 & D \\ \hline
\vphantom{\widehat{D}}\bar{D} & 0 
\end{array}
\end{array} \right) \left(\begin{array}{c|c}
0 & * \\ \hline
* & \begin{array}{c|c}
* & \boxbslash \\ \hline 
\boxbslash & *
\end{array}
\end{array} \right) \right) 
= \tr \left(\begin{array}{c|c}
0 & * \\ \hline
* & \begin{array}{c|c}
\boxbslash & *\\ \hline
* & \boxbslash 
\end{array}
\end{array} \right) = 0 \ . 
\end{equation} 
Here, $D$ is a placeholder for an arbitrary diagonal matrix, and
$\boxbslash$ for any off-diagonal matrix.

For a toroidal direction, $\mu \cong \chi^{l}_{(i)}$ ($i=1,2$), one needs the
explicit expressions for the matrices,
\begin{equation}
 \big(\Omega\big)^{\mu}_{ab}\Big|_{z=0} =-\frac{1}{2}
\left(\begin{array}{c|c}
0&0\\ \hline
0&\begin{array}{c|c}
0&\Delta^{\mu}\\ \hline 
\vphantom{\widehat{\Delta}}\bar{\Delta}^{\mu}&0\\
\end{array}
\end{array}\right) \ ,
\end{equation}
with
\begin{align}\label{def_Delta}
 \Delta_{mn}^{\mu} & =
\frac{k_{1}}{k_{1}+k_{2}}(\mathscr{G}^{+}_{l})_{mn}(k_2'\chi)
+\frac{k_2}{k_1+k_2}
(\mathscr{G}^{+}_{l})_{mn}(-k_1'\chi) \nonumber\\
& =\frac{2i\alpha^{n}_{l}}{k_1+k_2}\Big[k_1+k_2-
\big(k_1 e^{ik_2'(\chi^k\alpha^{n}_k)}
+k_2 e^{-ik_1'(\chi^k\alpha^{n}_k)}\Big)]\delta_{mn}\quad \text{for}\
\mu \cong \chi^{l}_{(1,2)}.
\end{align}
Comparing this result with the expressions~\eqref{expl_gammaplusbeta}
that we got for the constituent matrices $\hat{\gamma}+\hat{\beta}$ of
$\hat{g}+\hat{\omega}$, we find the relationship
\begin{equation}
\Delta^{\mu}_{mn} = \frac{i\alpha_{l}^{n}}{k_{1}+k_{2}}
(\hat{\gamma}+\hat{\beta})_{mn} + \mathcal{O} (z)\quad \text{for}\ \mu
\cong \chi^{l}_{(1,2)}.
\end{equation}
This is the crucial property that helps us to verify the equations of
motion~\eqref{EOMr}:
\begin{align}
 &\tr \Big[\big(\hat{g}+\hat{\omega}\big)^{-1}
\Omega^{\mu}\Big]\bigg|_{z=0}\nonumber \\
& \qquad = - \tr  \left[\left(\begin{array}{c|c}
\Big[\frac{1}{k_{1}'k_{2}' (k_{1}+k_{2})}\Big]\mathbbm{1}_r&0\\[1mm] \hline
0&\left( \begin{array}{c|c}
0&\hat{\gamma}+\hat{\beta}\\ \hline 
\overline{\hat{\gamma}+\hat{\beta}}^{\vphantom{\bar{a}}}&0\\
\end{array} \right)^{-1}
\end{array}\right) \left(\begin{array}{c|c}
0&0\\ \hline
0&\begin{array}{c|c}
0&\Delta^{\mu}\\ \hline 
\vphantom{\widehat{D}}\bar{\Delta}^{\mu}&0\\
\end{array}
\end{array}\right)\right]_{z=0}\nonumber \\ 
& \qquad = -\frac{1}{k_{1}+k_{2}}\tr \left[\left(\begin{array}{c|c}
\big( i\alpha^{n}_{l}\delta_{mn}\big)_{mn} & 0\\[1mm] \hline
0 & \big( -i\alpha^{n}_{l}\delta_{mn}\big)_{mn}
\end{array} \right) \right] \qquad (\mu \cong \chi^{l}_{(1,2)})\nonumber \\
& \qquad = 0 \ .
\end{align}
This finally shows that the generalised permutation brane given by the
embedding~\eqref{genpermbrane} and the boundary two-form
$\hat{\omega}$ stated in~\eqref{boundarytwoform} is a solution of the
DBI equations of motion.

\section{Further directions}

We have shown in this paper that the generalised permutation branes
are solutions of the DBI equations for products $G_{k_{1}}\times
G_{k_{2}}$ with an arbitrary compact simply connected, simple Lie
group $G$. A natural extension of this work would be to investigate
generalised permutation D-branes in products of coset models. As
mentioned in the introduction, such generalisations have been
formulated in the Landau-Ginzburg description of $N=2$ minimal models,
which can also be described as cosets $SU (2)/U (1)$, their geometric
interpretation is however unclear. Proposals for the geometries of
such D-branes in coset models $G/H$ have been made
in~\cite{Fredenhagen:2005an,Sarkissian:2006xp}, but it is questionable
whether any of these proposals is correct, as no successful DBI
analysis could be carried out so far. Another approach to find the
geometries of these branes in products like $SU (2)/U (1)\times SU
(2)/U (1)$ is by starting with generalised permutation branes on $SU
(2)\times SU (2)$, and then marginally perturb the $SU (2)$'s by
current-current deformations (see e.g.\cite{Giveon:1993ph}). At the
end-point of the deformation one expects a decoupling of one dimension
and the $SU (2)$'s reduce essentially to cosets $SU (2)/U (1)$ (see
e.g.\cite{Forste:2001gn}). In general it can happen that the branes
start to flow when the background is deformed (see
e.g.\cite{Fredenhagen:2006dn}), but it is conceivable that it is
possible to tune the deformations of the two factors such that the
generalised permutation brane does not flow. This is currently under
investigation.

\section*{Acknowledgements}

We thank Sebastian Krug and Rafa{\l} Suszek for interesting
and useful discussions.

\end{document}